\documentclass[twocolumn,showpacs,aps,prb,superscriptaddress,floatfix]{revtex4}
\usepackage[dvips]{graphicx}% Include figure files
\usepackage{dcolumn}% Align table columns on decimal point
\usepackage{bm}% bold math

\begin{document}

\title{Spin polarization and metallic behavior of a silicon two-dimensional electron system}
\author{Tohru Okamoto}
\affiliation{Department of Physics, University of Tokyo, Hongo, Bunkyo-ku, Tokyo
 113-0033, Japan}
\affiliation{Department of Physics, Gakushuin University, Mejiro, Toshima-ku, Tokyo
 171-8588, Japan}
\author{Mitsuaki Ooya}
\affiliation{Department of Physics, University of Tokyo, Hongo, Bunkyo-ku, Tokyo
 113-0033, Japan}
\author{Kunio Hosoya}
\affiliation{Department of Physics, Gakushuin University, Mejiro, Toshima-ku, Tokyo
 171-8588, Japan}
\author{Shinji Kawaji}
\affiliation{Department of Physics, Gakushuin University, Mejiro, Toshima-ku, Tokyo
 171-8588, Japan}

\date{\today}

\begin{abstract}
We have studied the magnetic and transport properties of an ultra-low-resistivity two-dimensional electron system in a Si/SiGe quantum well. The spin polarization increases linearly with the in-plane magnetic field and the enhancement of the spin susceptibility is consistent with that in Si-MOS structures. Temperature dependence of resistivity remains metallic even in strong magnetic fields where the spin degree of freedom is frozen out. We also found a magnetoresistance anisotropy with respect to an angle between the current and the in-plane magnetic field.
\end{abstract}
\pacs{71.30.+h, 73.40.Lq, 75.70.Cn}

\maketitle

Strongly correlated two-dimensional (2D) systems have attracted a great deal of attention in the last decade.
Metallic temperature dependence of resistivity $\rho (T)$ has been observed in a number of 2D systems where $r_s$ (the ratio of Coulomb interaction energy to Fermi energy) is large. \cite{Abrahams}
Although various theories have been proposed to explain the metallic behavior, its origin is still a subject of great debate.
Suppression of the metallic behavior by a strong magnetic field applied parallel to the 2D plane  has been reported for 2D electron systems in silicon metal-oxide-semiconductor (Si-MOS) structures \cite{Simonian,Mertes} and 2D hole systems in GaAs/AlGaAs heterostructures. \cite{Yoon,Tutuc-p,Gao}
Since the in-plane magnetic field $B_{||}$ does not couple to the 2D motion of carriers, the $B_{||}$-induced metal-insulator transition (MIT) is related to the spin of electrons (or holes).

The spin polarization in 2D Fermi liquid is given by $P=\mu_B g_v g_{\text {FL}} m_{\text {FL}} B_{||}/ 2 \pi \hbar^2 N_s$.
Here, $\mu_B$ is the Bohr magneton, $g_v$ is the valley degeneracy, and $N_s$ is the carrier concentration.
The effective $g$-factor $g_{\text {FL}}$ and the effective mass $m_{\text {FL}}$ are expected to be enhanced because of the interaction effect and larger than the band values of $g_b$ and $m_b$.
The enhancement factor $\alpha= g_{\text {FL}} m_{\text {FL}} / g_b m_b$ determined from Shubnikov-de Haas (SdH) oscillations was found to increase with $r_s$ in 2D electron systems in Si-MOS structures \cite{Okamoto99,Solidcom,Pudalov-mg,Shashkin} and a GaAs/AlGaAs heterostructure. \cite{Zhu}
The enhancement of the spin susceptibility leads to the reduction of the critical magnetic field $B_c$ for the full spin polarization ($P \rightarrow 1$).
Saturation of positive in-plane magnetoresistance (or sharp decrease in $d \rho / d B_{||}$) was associated with the onset of the full spin polarization at the reduced critical magnetic field. \cite{Tutuc-p,Okamoto99,Vitkalov,Tutuc-n}
Recently, Zhu {\textit {et al.}} \cite{Zhu} observed a nonlinear $B_{||}$-dependence of $P$ in a GaAs 2D electron system
 and explained the discrepancy between $\alpha$ determined from $B_c$ and from SdH oscillations at low magnetic fields.
It is not clear yet, however, whether this nonlinearity arises from intrinsic properties of 2D systems or
material-dependent properties, such as the spin-orbit interaction. \cite{Tutuc-n}
In silicon 2D electron systems, the Bychkov-Rashba spin-orbit parameter \cite{Rashba1,Rashba2}
is three orders of magnitude smaller than in 2D systems based on III-V semiconductors
and the band $g$-factor $g_b$ is 2.00. \cite{Wilamowski}
In Si-MOS structures, however, it is possible that disorder crucially changes the spin state of 2D electrons.
Pudalov {\textit {et al.}} demonstrated that
the in-plane magnetic field, at which the magnetoresistance saturates,
depends on the peak mobility of Si-MOS samples. \cite{Pudalov-dis}

In this paper, we report magnetotransport measurements on a silicon 2D electron system with a mobility two orders higher than that of high-mobility Si-MOS samples.
The in-plane magnetoresistance shows a kink corresponding to the onset of the spin polarization 
and an anisotropy with respect to an angle between the current and the magnetic field.
We obtain a linear relationship between $P$ and $B_{||}$ for $B_{||} \leq B_c$ and
$r_s$-dependence of $\alpha$ consistent with the results on Si-MOS samples.
Metallic temperature dependence of resistivity remains even in strong in-plane magnetic fields above $B_{c}$ in contrast to other systems where it is suppressed before the full spin polarization.

We use a Si/SiGe double heterostructure sample with a 20-nm-thick strained Si channel sandwiched between relaxed Si$_{0.8}$Ge$_{0.2}$ layers. \cite{Yutani1,Yutani2}
The electrons are provided by a Sb-$\delta$-doped layer 20~nm above the channel.
The 2D electron concentration $N_s$ can be controlled by varying bias voltage of a $p$-type Si (001) substrate 2.1~$\mu$m below the channel at 20~K and
determined from the Hall coefficient at low temperatures.
The 2D electron system has a high mobility of $ \mu = 66~{ \text m }^{2} / { \text {V}} { \text {s}} $ 
at $ N _ { s } = 2.2 \times 10 ^ { 15 }~{ \text m } ^ { -2 } $ 
(at zero substrate bias voltage) and $ T = 0.36~{ \text K } $.
Standard four-probe resistivity measurements were performed for a $600 \times 50~\mu {\textrm m}^2$ Hall bar sample mounted on a rotatory stage in a pumped $^3$He refrigerator together with a GaAs Hall generator and resistance thermometers.

\begin{figure}
\includegraphics[width=8cm]{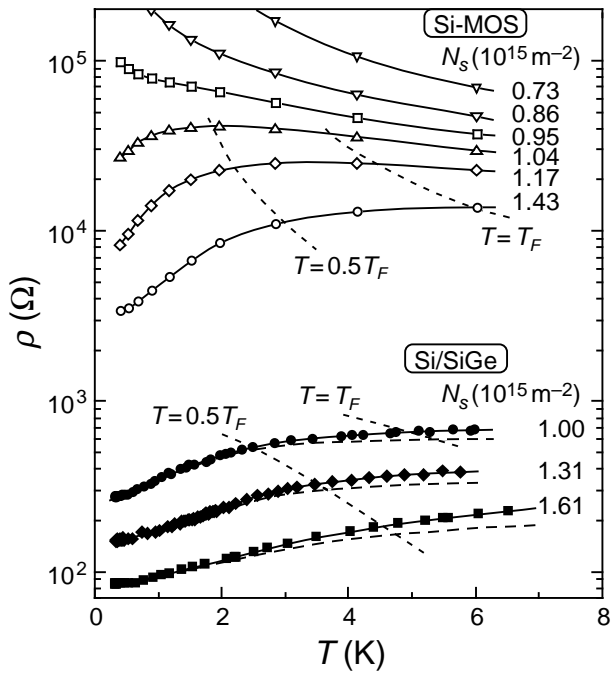}
\caption{\label{fig1} 
Temperature dependence of resistivity at $B=0$.
The closed symbols are for the Si/SiGe sample and
the open symbols for a Si-MOS sample used in Ref.~\protect\onlinecite{Okamoto99}.
The dotted lines mark $T=0.5T_F$ or $T=T_F$.
The contribution of phonon-scattering to $\rho$ in the Si/SiGe sample is calculated based on Ref.~\protect\onlinecite{Paul} and
subtracted from the experimental data (dashed lines).
}
\end{figure}

In Fig.~\ref{fig1}, we show the temperature dependence of resistivity in the Si/SiGe sample and a Si-MOS sample.
The latter has a peak mobility of $\mu_{\text {peak}} = 2.4~{ \text m }^{2} / { \text {V}} { \text {s}} $ and
exhibits an apparent MIT. \cite{Okamoto99}
The Fermi temperature is given by 
$T_{F} = 2 \pi \hbar^2 N_{s} / g_s g_v k_B m_{\text {FL}}$,
 where we have the spin degeneracy $g_s = 2$ at $B = 0$ and
$g_v = 2$ for the (001) silicon 2D electron systems.
The effective mass $m_{\text {FL}}$ enhanced from
the band mass of $m_b=0.19 m_e$ (Ref.~\onlinecite{Rieger}) is obtained from Ref.~\onlinecite{Pudalov-mg}.
In the Si/SiGe sample, the contribution of phonon scattering to $\rho$ (Ref.~\onlinecite{Paul}) is not negligible at high temperatures,
 but it is not important in the low temperature region of $T \lesssim 0.5 T_{F}$ where the strong metallic temperature dependence is observed.
Although the resistivity drop at low temperatures in the Si/SiGe sample
is rather weaker than that in the Si-MOS sample observed in the vicinity of the MIT,
the overall behavior of the $T$-dependence curves (or $T/T_F$-dependences) are similar.
The common feature of $T$-dependence of $\rho$ suggests that the origin of the metallic behavior is the same in these two types of silicon 2D electron systems having quite different structures and mobilities.

\begin{figure}
\includegraphics[width=8cm]{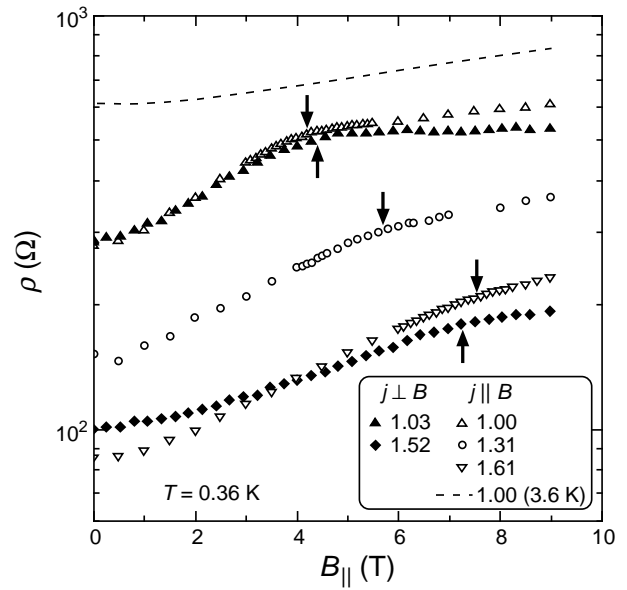}
\caption{\label{fig2} 
Resistivity at 0.36~K (except the dashed line at 3.6~K) as a function of the in-plane magnetic field
for different electron concentrations $N_s (10^{15} {\text m}^{-2})$.
The closed symbols are for the current orientation of $j \perp B$
and the open symbols for $j \parallel B$.
The arrows indicate the critical magnetic field for the full spin polarization.
}
\end{figure}

Figure~\ref{fig2} shows the in-plane magnetoresistance in the Si/SiGe sample.
We observe an abrupt change in the slope of the $\rho$ vs $B_{||}$ curve indicated by arrows.
As will be discussed later, it corresponds to the onset of the full spin polarization of the 2D electrons.
This kink is smeared out at high temperatures.
The magnetoresistance depends on the current orientation with respect to the in-plane magnetic field and it is larger for $j \parallel B$ than for $j \perp B$.
A similar anisotropy was also found in a Si-MOS sample \cite{Pudalov-dis}
although it was smaller than that in the Si/SiGe sample.
Besides the contribution of the spin polarization that makes the kink, we should take account of
the contribution of the orbital effect to the in-plane magnetoresistance 
owing to the finite thickness of the 2D systems.
The orbital effect is expected to be stronger
in the wide quantum well ($ = 20$~nm) in the Si/SiGe sample
than in the narrow channel ($< 10$~nm) in Si-MOS structures.
We consider that the anisotropy arises from the orbital effect.
In the classical view, on the other hand, 
a magnetic field does not affect the current flowing parallel to it and
we simply expect smaller magnetoresistance for $j \parallel B$.
The opposite anisotropy observed in silicon 2D systems
 is an open question.

\begin{figure}
\includegraphics[width=8cm]{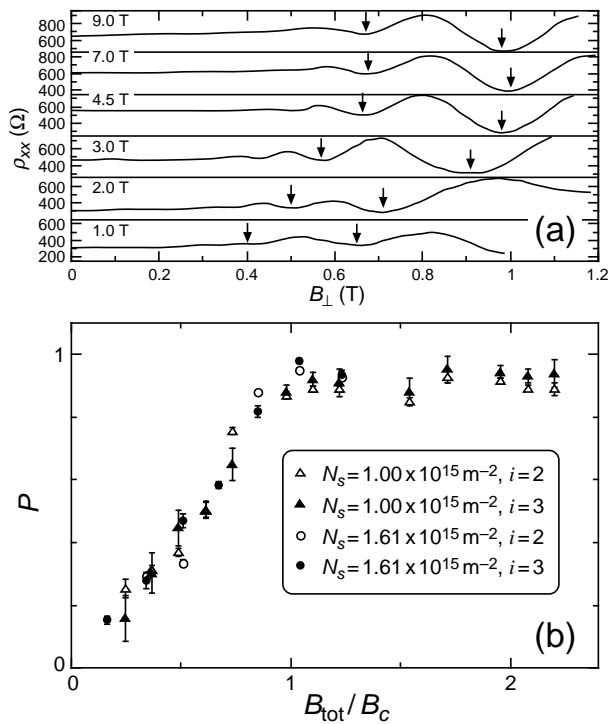}
\caption{\label{fig3} 
(a) Shubnikov-de Haas oscillations at $T=0.36$~K and $N_s =1.00 \times 10^{15}$~m${}^{-2}$
for different $B_{\text {tot}}$.
(b) The spin polarization obtained from $B_\perp$ at the $\rho_{xx}$ minima.
}
\end{figure}

The spin polarization $P$ is determined from SdH oscillations
as a function of the total strength $B_{\text {tot}}$ of the magnetic field.
\cite{Okamoto99,Tutuc-p,Tutuc-n,Okamoto00}
By rotating the sample in a constant magnetic field,
we introduce the perpendicular component $B_\perp$
and observe an oscillation of the diagonal resistivity $\rho_{xx}$
as shown in Fig.~\ref{fig3}~(a).
The $\rho_{xx}$ minima indicated by arrows
shift toward higher-$B_\perp$ side as $B_{\text {tot}}$ increases.
This feature is associated with the concentration $N_{\uparrow}$ of spin-up electrons.
The observed $\rho_{xx}$ minima correspond to 
$B_\perp = N_\uparrow h/ i g_{v} e$ with $i = 2$ or 3.
In Fig.~\ref{fig3}~(b), we show the spin polarization $P= 2 N_\uparrow /N_s - 1$
as a function of $B_{\text {tot}}$.
No systematic differences are found between the data for $i=2$ and $i=3$.
We believe that $B_\perp$ used here is small enough and 
$P$ is determined in the limit of $B_\perp =0$ ($B_{\text {tot}}=B_{||}$) .
It is confirmed that the increase in $P$ saturates at $B_c$ determined from
the magnetoresistance curve shown in Fig.~\ref{fig2}.
The observed linear relationship of $P$ with $B_{||}$
for $B_{||} \leq B_c$ demonstrates
that $\alpha=g_{\text {FL}} m_{\text {FL}} / g_{b} m_{b}$ does not depend on $P$
in contrast to the case of a GaAs 2D electron system. \cite{Zhu}

\begin{figure}
\includegraphics[width=8cm]{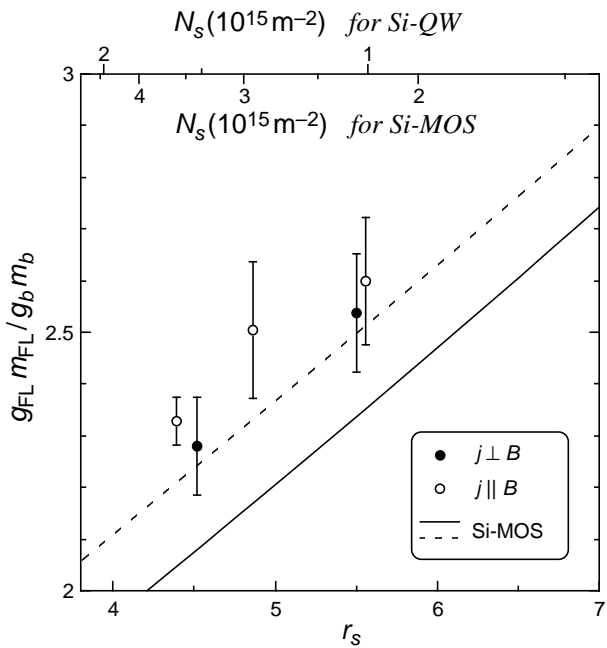}
\caption{\label{fig4} 
$r_{s}$ dependence of $\alpha= g_{\text {FL}} m_{\text {FL}} / g_{b} m_{b}$. 
The closed and open circles are determined 
from $B_c$ in the magnetoresistance shown in Fig.~\protect\ref{fig2}
using $B_c = 2 \pi \hbar^2 N_s / \mu_B g_v g_{\text {FL}} m_{\text {FL}} $.
The solid line shows the experimental data 
on Si-MOS structures
with $\kappa=7.7$. \protect\cite{Okamoto99,Pudalov-mg}
$N_s$ is also indicated for the silicon quantum well ($\kappa = 11.5$)
and for Si-MOS structures ($\kappa= 7.7$).
The dashed line is after the correction of $\kappa$ (see text).
}
\end{figure}

In Fig.~\ref{fig4}, $\alpha$ determined from $B_c$ is shown as 
a function of $r_s = \pi^{1/2} (e/h)^2 (m_b / \kappa \epsilon_{0}) N_s{}^{-1/2}$.
The relative dielectric constant $\kappa=11.5$ is used for the silicon quantum well. \cite{kappaSiGe}
In Refs.~\onlinecite{Okamoto99} and \onlinecite{Pudalov-mg}, $\alpha$ in Si-MOS structures was obtained from SdH oscillations and $r_{s}$ was calculated from $N_s$ using $\kappa=7.7$,
average relative dielectric constant of silicon and SiO$_2$. \cite{Ando}
$\alpha(r_s)$ in this work is almost consistent with Refs.~\onlinecite{Okamoto99} and \onlinecite{Pudalov-mg}
but slightly ($\sim 10~\%$) higher.
This difference may arise from an overestimation of $r_s$ 
in Si-MOS structures.
The average distance of electrons from the Si/SiO$_2$ interface is calculated to be $z_0 \approx 3$~nm in the range of $N_{s} =2 \sim 4 \times 10^{15} {\text m}^{-2}$. \cite{Ando}
It is smaller than, but comparable to the average distance between electrons $(\pi N_s)^{-1/2} \sim 10$~nm.
Thus the relative dielectric constant should be effectively larger than 7.7.
We estimate the effective value of $\kappa$
from the calculation of the Coulomb force between electrons located at the distance $z_0$ away from the interface and separated each other by $(\pi N_s)^{-1/2}$.
This correction leads to smaller $r_s$ and better agreement of $\alpha (r_s)$ 
with the present data for the Si/SiGe sample. 

\begin{figure}
\includegraphics[width=8cm]{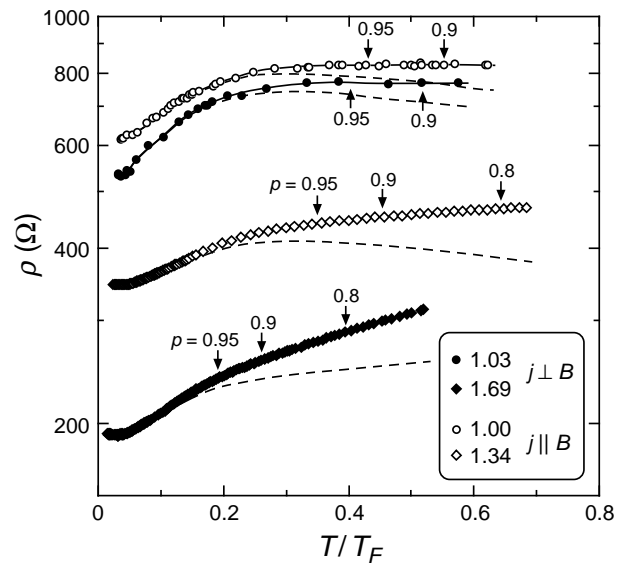}
\caption{\label{fig5} 
$\rho$ vs $T/T_F$ in in-plane magnetic fields above $B_{c}$:
$B_{||}=9$~T for $N_{s} =1.00$ and $1.03 \times 10^{15} {\text m}^{-2}$, 
and $B_{||}=11$~T for $N_{s} =1.34$ and $1.69 \times 10^{15} {\text m}^{-2}$.
The spin polarization $P$ is indicated for each $N_s$.
The contribution of phonon-scattering is calculated based on Ref.~\protect\onlinecite{Paul} and
subtracted from the experimental data (dashed lines).
}
\end{figure}

\begin{figure}
\includegraphics[width=8cm]{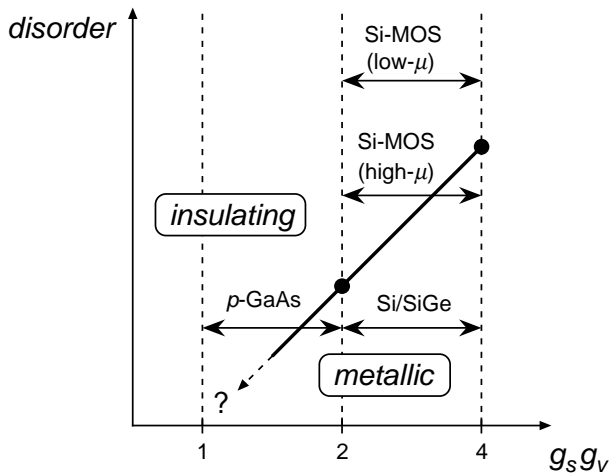}
\caption{\label{fig6} 
Schematic phase diagram for $T$-dependence of $\rho$ in low-resistivity and strongly-correlated 2D systems.
}
\end{figure}

Figure~\ref{fig5} shows the dependence of $\rho$ on $T/T_F$
in strong in-plane magnetic fields above $B_{c}$.
$T_{F}$ is given for the full spin polarization ($g_{s}=1$)
and $P$ is calculated from the Fermi distribution function.
The apparent metallic behavior is observed at low temperatures,
where the spin of electrons is almost polarized.
It is in contrast to the results on Si-MOS structures \cite{Simonian,Mertes} and $p$-type GaAs/AlGaAs heterostructures \cite{Yoon,Tutuc-p,Gao}
where the metallic behavior disappears before the full spin polarization
even for resistivity much lower than the critical resistivity ($\sim h/e^2$) 
at the MIT in the absence of a magnetic field.
In Fig.~\ref{fig6}, we propose a schematic phase diagram for $T$-dependence of $\rho$ in low-resistivity ($\rho \lesssim h/e^2$) and strongly-correlated ($r_s \gg 1$) 2D systems.
We consider that the internal degree of freedom is essential for the metallic behavior.
In the case of Si-MOS structures, the metallic behavior is observed at $B=0$
in samples having a high peak mobility 
($\mu_{\text {peak}} \gtrsim 2~{ \text m }^{2} / { \text {V}} { \text {s}} $). \cite{Abrahams}
The strong in-plane magnetic field changes the degeneracy factor $g_s g_v$ from 4 to 2
and suppresses the metallic behavior.
This indicates that the critical level of disorder is lower for $g_s g_v = 2$ than
that for $g_s g_v = 4$.
The metallic behavior with $g_s g_v=2$ can be observed in the heterostructure systems
with much higher mobility than that of Si-MOS samples.
We have $g_s=2$ and $g_v=1$ in GaAs 2D hole systems at $B=0$,
and $g_s=1$ and $g_v=2$ in silicon 2D electron systems at $B_{||}> B_c$. \cite{valley}
Since the $B_{||}$-induced MIT is observed even in high-mobility
GaAs 2D hole systems, \cite{Tutuc-p}
the critical level of disorder, if exists, is expected to be very low for $g_s g_v =1$.

In summary, we have studied the spin polarization and $T$-dependence of resistivity
in an ultra-high-mobility Si/SiGe heterostructure sample.
The spin polarization increases linearly with the in-plane magnetic field
in contrast to the case of a GaAs 2D electrons system
and the $r_s$-dependence of the spin susceptibility
is consistent with the previous measurements on Si-MOS samples.
We observed apparent metallic $T$-dependence of $\rho$ for $B_\parallel >B_c$
in contrast to other systems where it disappears before the full spin polarization.
We consider that this is owing to high mobility (low disorder)
and the valley degree of freedom in the Si/SiGe sample.
A resistance anisotropy with respect to an angle between the current and the in-plane magnetic field is also found.

We thank Professor Y. Shiraki for providing us with the Si/SiGe sample.
This work is supported in part by Grants-in-Aid for Scientific Research from the Ministry of Education, Science, Sports, and Culture, Japan.


\begin{references}
\bibitem{kappaSiGe} We estimate the effective value of $\kappa$ in the 20~nm-thick silicon quantum well in the Si/Si$_{0.8}$Ge$_{0.2}$ double heterostructure is close to that in bulk silicon.
\bibitem{valley} We estimate the valley splitting energy to be less than 0.4~K from SdH
measurements.
\end{references}
\end{document}